\documentclass[a4paper]{article}
\usepackage[left=2cm, right=2cm, top=2cm,bottom=3.5cm]{geometry}
\usepackage[cp1251]{inputenc}
\usepackage[T2A]{fontenc} 
\usepackage[pdftex]{color}
\usepackage[english]{babel}
\usepackage[active]{srcltx}
\usepackage{multicol,pgf,tikz,pgfplots,textcomp, eufrak, euscript, latexsym,blindtext}
\usepackage{amssymb,amsfonts,amsmath,stmaryrd,mathtext,cite,enumerate,float,graphicx,amstext, longtable,amsthm,url, centernot,caption}
\usepackage{indentfirst}
\usepackage{fancybox, float}
\usepackage{longtable}
\usepackage{eurosym}
\usepackage{multirow}
\usepackage{calrsfs}
\usepackage{authblk}
\DeclareMathAlphabet{\pazocal}{OMS}{zplm}{m}{n}
\usetikzlibrary{arrows}

\theoremstyle{plain}
\newtheorem{theorem}{Theorem}[section]

\theoremstyle{definition}

\newtheorem{definition}[theorem]{Definition}

\usepackage[bookmarks=true,hyperindex,pdftex,colorlinks,citecolor=red,linkcolor=black,urlcolor=black,linkbordercolor=black]
{hyperref}

 \numberwithin{equation}{section}

\renewcommand{\frac}{\dfrac}

\renewcommand{\geq}{\geqslant}

\newcommand{\R}{{\mathbb{R}}}
\newcommand{\erf}{\operatorname {erf}}

\newcommand{\N}{{\mathbb{N}}}

\renewcommand{\geq}{\geqslant}

\newcommand{\bea}{\begin{eqnarray*}}
\newcommand{\eea}{\end{eqnarray*}}
\newcommand{\beq}{\begin{equation}}
\newcommand{\eeq}{\end{equation}}
\newcommand{\begsta}{\begin{statements}}
\def\endsta{\end{statements}}
\newcommand{\begaeq}{\begin{aequivalenz}}
\def\endaeq{\end{aequivalenz}}

\title{Efficient representation of supply and demand curves on day-ahead electricity markets}
\author{Mariia Soloviova \thanks{\noindent Department of Mathematics ``Tullio Levi Civita'', University of Padua; soloviov@math.unipd.it}\qquad
Tiziano Vargiolu  {\footnote{Department of Mathematics ``Tullio Levi Civita'', University of Padua and Interdepartmental Centre for Energy Economics and Technology ``Giorgio Levi-Cases'', University of Padua; vargiolu@math.unipd.it}} }
\date{  
}

\begin{document}

\maketitle
\large
\begin{abstract}
Our paper aims to model supply and demand curves of electricity day-ahead auction in a parsimonious way. 
Our main task  is to build an appropriate algorithm to present the information about electricity prices and demands with far less parameters than the original one.
We represent each curve using mesh-free interpolation techniques based on radial basis function approximation.
We describe results of this method for the day-ahead IPEX  spot price of Italy.

\end{abstract}

\section{Introduction}
\begin{figure}[H]
\begin{minipage}[b]{.4\textwidth}
\centering
\includegraphics[width=1\textwidth]{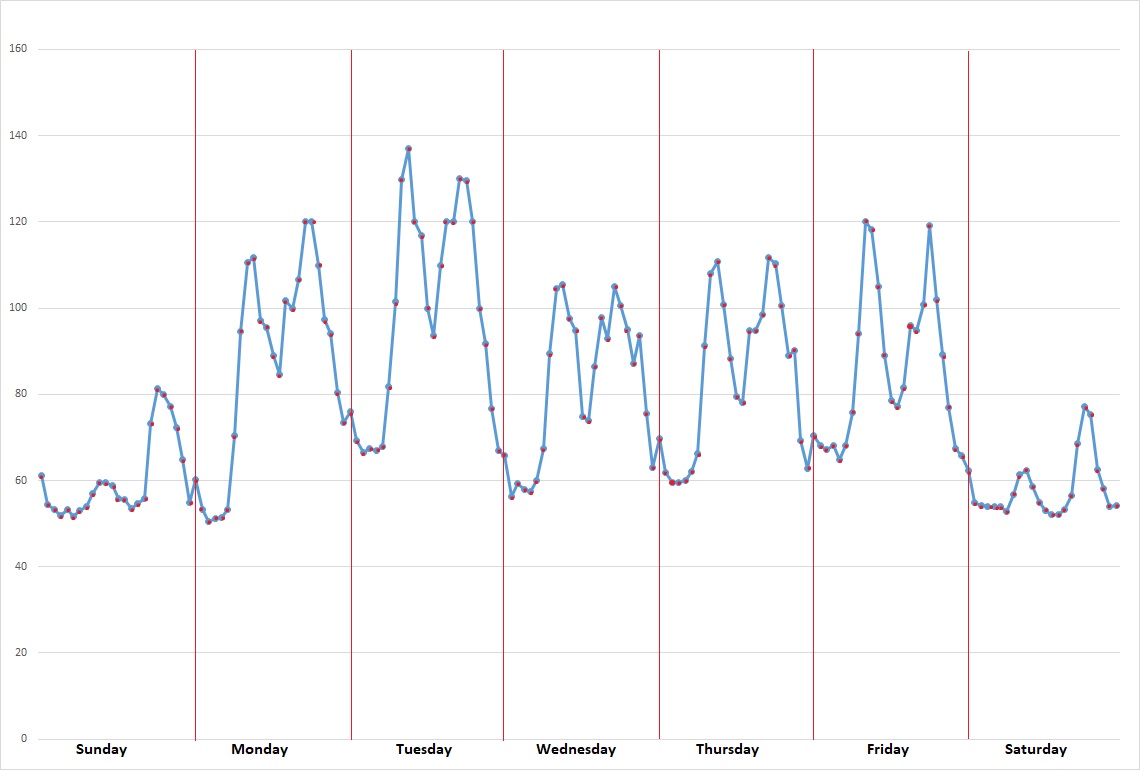}
\caption{Electricity equilibrium  prices during a week}
\label{fig_price_week}
\end{minipage}
\hfill
\begin{minipage}[b]{.4\textwidth}
\centering
\includegraphics[width=1\textwidth]{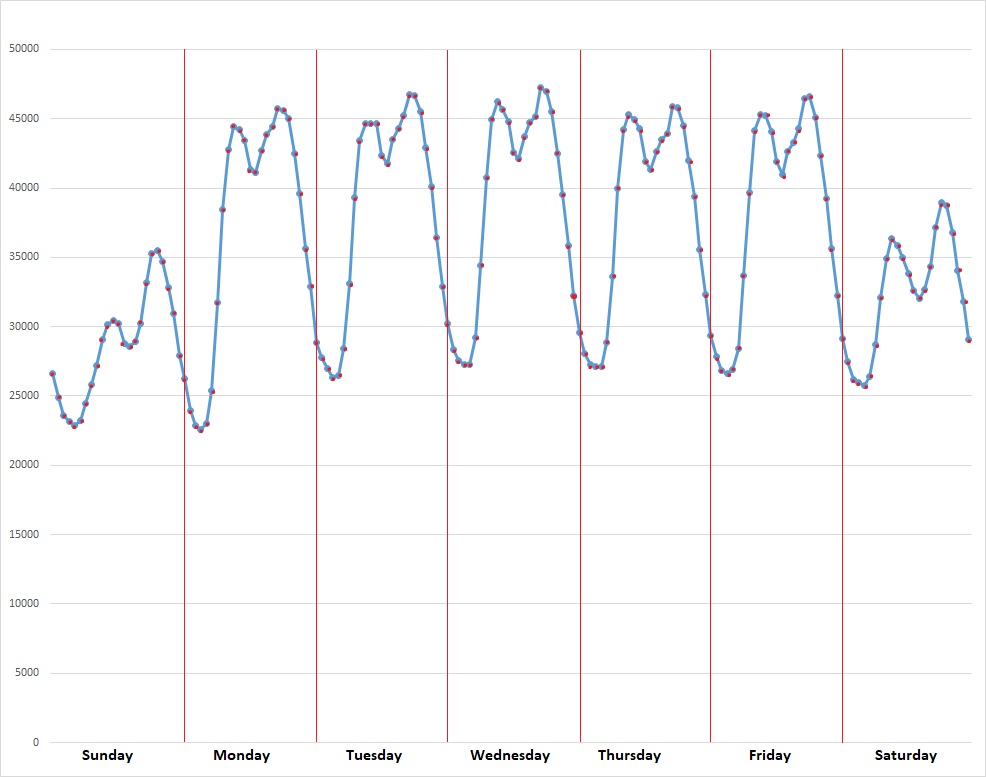}
\caption{Electricity equilibrium quantities during a week} 
\label{fig_quantity_week}
\end{minipage}
\end{figure}

Accurate modeling and forecasting electricity demand and prices are very important issues for decision making in deregulated electricity markets. Different techniques were developed
to describe and forecast the dynamics of electricity load. Short term forecast proved to be very challenging task due to these specific features. Figure \ref{fig_price_week} and  \ref{fig_quantity_week} demonstrate changing of electricity equilibrium  price and  quantity during one week. Functional data analysis is extensively used in other fields of science, but it  has been little explored in the electricity market setting.

We consider the Italian electricity market (IPEX). IPEX consists of different markets, including a day-ahead market. The day-ahead market is managed by Gestore del Mercato Elettrico where prices and demand are determined the day before the delivery. Supply and demand curves on day-ahead electricity markets are the results of thousands of bid and ask entries in the day-ahead auction, this for all the 24 hours. In principle, it would be possible to represent, and forecast, these curves by taking into account each production and each consumption unit as a separate time series, and then joining these together to construct the final curves, and thus the resulting price. However, the huge number of these units makes this naive strategy infeasible, unless one has extremely high computing capacity with complex machine learning algorithms available. 

In this paper, we are going to present a more parsimonious approach. In fact, the idea is to represent each curve using non-parametric mesh-free interpolation techniques, so that we can obtain an approximation of the original curve with far less parameters than the original one. The original curve, in fact, in principle depends on about hundreds of parameters and is obtained as follow.

\begin{figure}[h]
\centering
\includegraphics[width=1\textwidth]{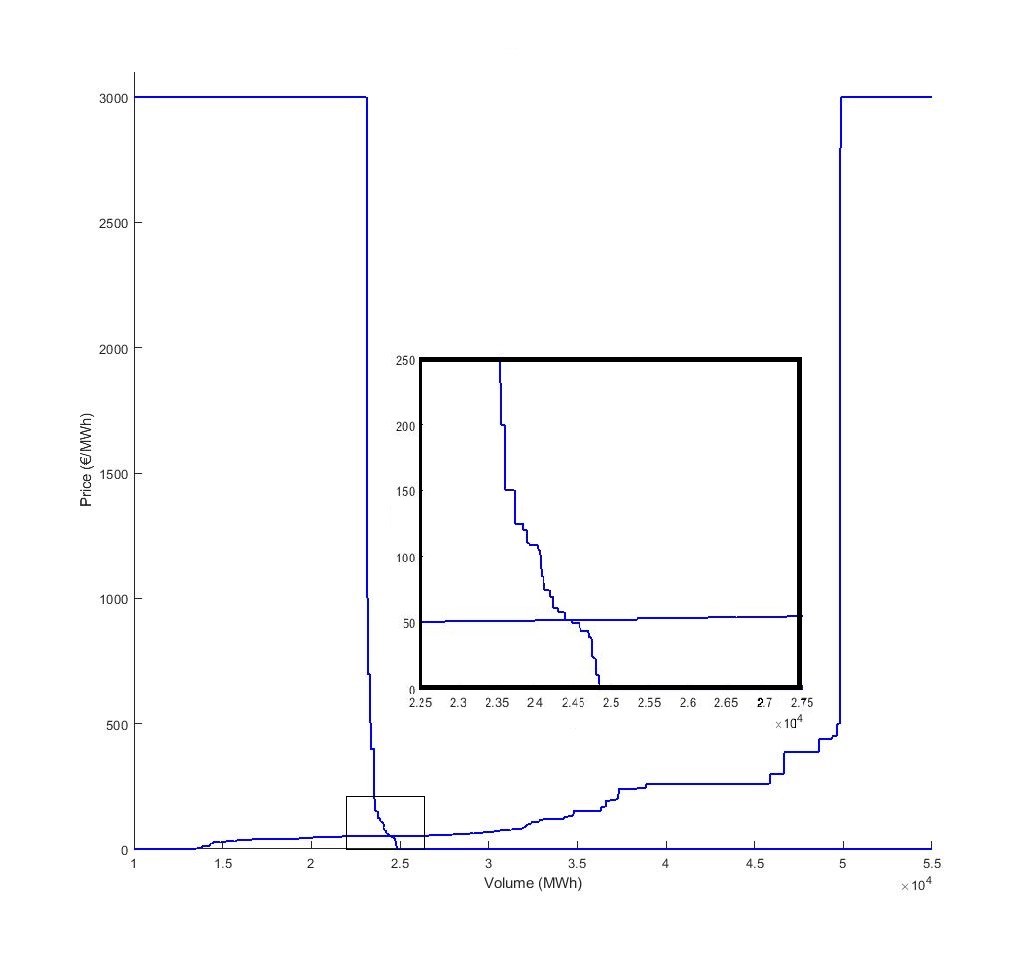}  
\caption{The market equilibrium point}
\label{Fig_equilibr_point}
\end{figure}

The   producers   submit offers   where they specify the quantities and the minimum  price at which they are willing to sell. The  demanders submit  bids where they specify
the quantities and the maximum price at which they are willing to buy. They are then aggregated by an independent system operator (ISO) in order to construct the supply and demand curves. 
Once the offers and bids are received by the ISO, supply and demand curves are established
by summing up individual supply and demand schedules. In the case of demand, the first
step is to replace ''zero prices`` bids by the market maximum price (for Italian electricity market, the market
maximum price is 3000 Euro) without changing the corresponding quantities. After this
replacement, the bids are sorted from the highest to the lowest with respect to prices. The
corresponding value of the quantities is obtained by cumulating each single demand bid.
For supply curve, in contrast, the offers are sorted from the lowest to the highest with respect to prices and the corresponding value of the quantities is obtained by cumulating each
single supply offer.  The market equilibrium is
the point where both curves intersect each other and the price balances supply and demand
schedules (see, e.g. Figure \ref{Fig_equilibr_point}). This point determines the market clearing price and the traded quantity. Accepted offers and bids are those that fall to the left of the intersection of the two curves, and all of them are exchanged at the resulted price. 



In the beginning of the 2000s the amount of papers focused on electricity price forecasting started to increase dramatically. A great variety of methods and models occurred during last twenty years. Weron \cite{Weron} (2014) made an overview of the existing literature on electricity price forecasting and divided  electricity price models into five different groups: multi-agent, fundamental, reduced-form, statistical and
computational intelligence models. A review of probabilistic forecasting was done in \cite{Weron2} (2018) by Weron and Nowotarski. Most models have in common that they focus on the price itself or related time series. In such a way these models does not take into account the underlying mechanic which determines the price process -- the intersection between the part of the electricity supply and demand.

Some of the recent approaches try to to analyse the real offered volumes for selling and purchasing electricity. This commonly leads to a problem of a large amount of data and, therefore, high complexity. In particular, Eichler, Sollie, Tuerk in 2012 \cite{Eichler}  investigated a new approach that exploits information available in the supply and demand curves for the German day-ahead market. They  proposed the idea that the form of the supply and demand curves or, more precisely, the
spread between supply and demand, reflects the risk of extreme price fluctuations. They
utilize the curves to model a scaled supply and demand spread using
an autoregressive time series model in order to construct a flexible model adapted to changing
market conditions. Furthermore, Aneiros, Vilar, Cao, San Roque in 2013 \cite{Aneiros} dealt with the prediction of residual demand curve in elecricity spot market using two functional models. They  tested this method as a tool for optimizing bidding strategies for the Spanish day-ahead market.   Then Ziel and Steinert in 2016 \cite{Ziel1} proposed a model for the German European Power Exchange (EPEX) market, which considers all the supply and demand information of the system and discusses the effects of the changes in supply and demand. Their idea was to fill the gap between research done in time-series
analysis, where the structure of the market is usually left out, and the research done in structural analysis, where empirical data is utilized very rarely and even less thoroughly. They provided deep insight on the
bidding behavior of market participants. They also showed that incorporating the sale and purchase data yields promising results for forecasting the likelihood of extreme price
events. In 2016  Shah \cite{Tesi} also considered the idea of modeling the daily supply and demand curves, predicting them and finding the intersection of the predicted curves in order to find the predicted market clearing price and volume. He used the functional approach, namely, B-spline approximation, to convert the resulted piece-wise constant curves into smooth functions. 

As far as we know,  non-parametric mesh-free interpolation techniques were never considered for the problem of modeling the daily supply and demand curves. 







We  are going to use a relatively new modeling technique based on functional data analysis for demand and price prediction.  The first task for this purpose is to make an appropriate algorithm to present the information about electricity prices and demands, in particular to approximate a monotone piecewise constant  function.

We want to make an appropriate algorithm to present this information, in particular, to approximate a monotone piecewise constant  function. Accuracy of the approximation and running time are very important for us. As we already said, the basic novelty of our problem is that we are going to present the information about electricity prices and demands using functional data analysis approach. 
The main idea behind functional data analysis  is, instead of considering a collection of data points, to consider the  data as a single structured object. This  allows to use
additional information contained in the functional structure of the data.
Once the data are converted to functional form, it can be evaluated at all values over some interval.




The most promising technique 
to do so is the use of (integrals of) Radial Basis Functions, which 
are been used in several other applications (image reconstruction, medical 
imaging, geology, etc.) and allow a very flexible adaptation of the 
interpolating curves to real data. The use of radial basis functions have attracted increasing attention in recent years as an elegant scheme for high-dimensional scattered data approximation, an accepted method for machine learning, one of the foundations of meshfree methods and so on. The initial motivation for RBF  methods came from geodesy, mapping, and meteorology. RBF methods were first studied by Roland Hardy, an Iowa State geodesist, in 1968, when he developed one of the first effective methods for the interpolation
of scattered data.   Later in 1986 Charles Micchelli, an IBM mathematician, developed the theory behind the multiquadric method. Micchelli  made the connection between scattered data interpolation and positive definite functions \cite{Micchelli}.  RBF methods are now considered an effective way  to solve partial differential equations, to represent topographical surfaces as well as other intricate three-dimensional shapes, having been successfully applied in such diverse areas as climate modeling, facial recognition, topographical map production, auto and aircraft design, ocean floor mapping, and medical imaging (see, for example,\cite{Bozzini},\cite{Fasshauer2011}, \cite{Emma1}). Now RBF methods are an active area of mathematical research, as many open questions still remain.  We will present different techniques for this interpolation, with their advantages and drawbacks, and with an 
application to the Italian day-ahead market.

The paper is organized as follow. Section 2 describes the theoretical background, namely, mesh-free interpolation techniques based on radial basis function approximation.
Section 3 presents the database from the
Italian electricity market.
Section 4 is devoted to a short description of the numerical schemes and to the  analysis of the results.
Section 5 concludes the paper.
 


\section{Meshless approximation}

Let us briefly notice some features of supply and demand curves that are relevant for our modeling:
\begin{itemize}

\item By construction, the curves are monotone. 

\item The values attained by the supply curve are roughly clustered around {\bf layers}, corresponding to different 
production technologies.  In Italy they are non-dispatchable renewables, gas, coal, hydro, oil. 

\item The fact that renewables are the first ones make the supply curve intrinsically "meshless". 

\item Demand is much more inelastic than supply.
\end{itemize}
So, we are dealing with a scattered data interpolation problem. We have a large amount of points (each point represents price and amount of electricity) that we want to approximate. We can formalize this problem as follows.

Given a set of $N$ distinct \textit{data points}  $X_N=\{x_i: i=1, 2, \ldots, N\}$ arbitrarily distributed on a domain $\Omega\subset \R$ and a set of \textit{data values} (or function values) $Y_N=\{y_i: i=1, 2,\ldots, N\}\subset \R$, the data interpolation problem consists in finding a function $s_f: \Omega \rightarrow \R$ such that
\begin{equation}\label{interpol_equation}
s_f(x_i)=y_i,\, i=1,\ldots,N.
\end{equation} 

Let us recall briefly the most popular methods for the interpolation problem. Polynomial interpolation   is the interpolation of a given data set by the polynomial of lowest possible degree that passes through the points of the dataset.  For
given data sites $X_N$ and function values $Y_N$ there exists exactly one polynomial $p\in \pi_{N-1}(\R)$  that interpolates the data at the data sites. Therefore the space $\pi_{N-1}(\R)$ depends neither on the
data sites nor on the function values but only on the number of points.

\begin{figure}[h!]
\caption{Approximation of supply curve with polynomials}
\centering
\includegraphics[width=1\linewidth]{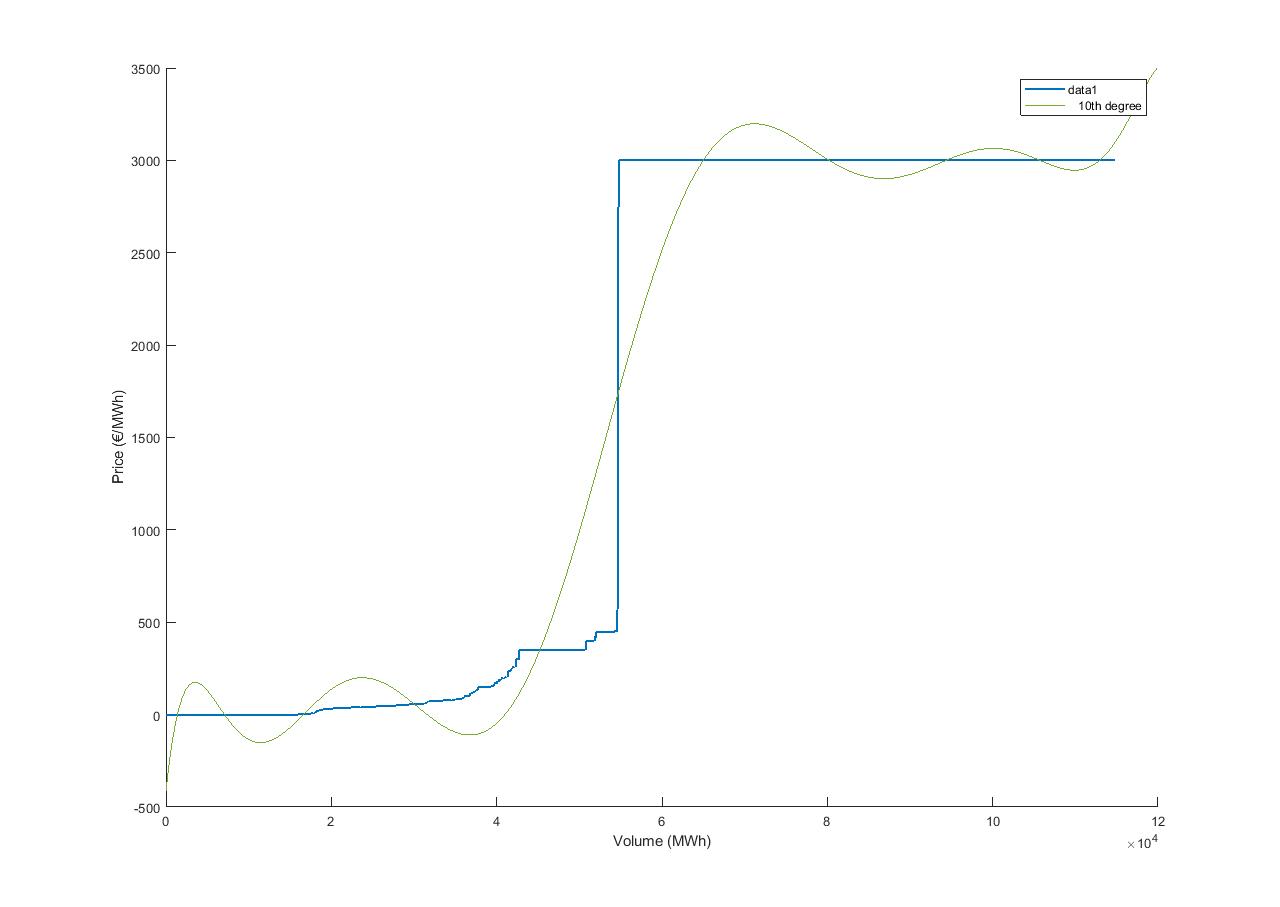}
\label{PlinomAppr}
\end{figure}

Runge's phenomenon (1901) shows that for high values of $N$, the interpolation polynomial may oscillate wildly between the data points.  Besides, the polynomial interpolation does not guarantee of monotonicity of the curves (see Figure \ref{PlinomAppr}).

It is a well-established fact that a large
data set is better dealt with splines than with polynomials. An aspect to notice in contrast
to polynomials is that the accuracy of the interpolation process using splines is not based on the
polynomial degree but on the spacing of the data sites. In particular, cubic splines are widely used to fit a smooth continuous function through discrete data. However, spline interpolation requires a mesh.

Notice that for all methods, the interpolant $s_f$ is expressed as a linear
combination of some basis functions $B_i$ , i.e. $$s_f(t)= \displaystyle \sum_{k=1}^d c_k B_k(t).$$
The basis functions in e.g. polynomial interpolation does not depend on the data points.  Another approach is to use a basis which depends on the data points.



One simple way to solve problem \eqref{interpol_equation} is to choose a fixed function $\phi:\R\rightarrow \R$
 and to form the interpolant as
$$s_f(x)= \displaystyle \sum_{i=1}^N \alpha_i \phi(\|x-x_i\|),$$
where the coefficients $\alpha_i$ are determined by the interpolation conditions
$s_f(x_i)=y_i$. Therefore,  the scattered data interpolation problem leads
to the solution of a linear system
\begin{equation*}
A\alpha=y, \text{ where } A_{i,j}=\phi(|x_i-x_j|).
\end{equation*}
 The solution of the system requires
that the matrix $A$ is non-singular. It is enough to know in advance that the matrix is positive definite (see \cite{Wendland} for more details).  Let us recall the definition of strictly positive definite function.

\begin{definition}\label{pos_def}
A real-valued function $\Phi:\R\longrightarrow \R$ is called
\textit{ positive semi-definite} if , for all $m\in \N$ and for any set of pairwise distinct points
$x_1, x_2, \ldots,x_m$, the $m\times m$ matrix
$$A=\left(\Phi(x_i-x_j)\right)_{i, j=1}^m$$
is positive semi-definite, i.e. for every column vector $z$ of $m$ real numbers the scalar $z^T A z\geq 0$. The function $\Phi:\R\longrightarrow \R$ is called (strictly)
\textit{ positive  definite} if the  matrix $A$ is positive definite, i.e. for every non-zero column vector $z$ of $m$ real numbers the scalar $z^T A z> 0$.
\end{definition}

The most important property of positive semi-definite matrices is that their eigenvalues are positive and so is its determinant. 

 A  radial function is a real-valued positive semi-definite function whose value depends only on the distance from the center $\mathbf {c}$. One useful characterization for positive semi-definite univariate function was given by Schoenberg in 1938 in the terms of completely monotone functions: a continuous function $\phi:[0, \infty)\rightarrow \R $ is positive semi-definite if and only if $\phi\in C^\infty(0, \infty)$ and $(-1)^k\phi^{(k)}(r)\geq 0$ for all $r\geq 0$, for $k=0, 1,\ldots$.

 Some standard radial basis functions are
 \begin{itemize}
 \item ${\displaystyle \phi (r)=e^{-(\varepsilon r)^{2}}} $ (Gaussian),
 \item  $\phi (r)=e^{-\varepsilon r}(\varepsilon r+1)$ (Mat\'{e}rn), 
 \item $\phi (r)=(1-\varepsilon r)^4_+(4\varepsilon r+1)$ (Wendland),
\end{itemize}   
where $\varepsilon>0$ denote a shape parameter, $r=\|x\|_2$.



The idea of meshless approximation with radial basis functions is to find an  approximant of $f$ in the following form:
$$ s_f(x) := \sum_{i=1}^N \alpha_i \phi(\|x - x_i\|) $$
where: 
\begin{itemize}
\item the coefficients $\alpha_i$ and the {\bf centers} $x_i$ are to be chosen so that the interpolant is as near as possible as the original function $f$; 
\item $\phi: \R \to \R$ is a  {\bf radial basis function} (RBF).
\end{itemize} 

Notice that the radial basis function $\phi \geq 0$, with $\alpha_i \geq 0$, so
$$ \sum_{i=1}^M \alpha_i \phi(\|x - x_i\|) \geq 0. $$ 
As we need to approximate piecewise constant monotone  function from $[0, M]$ to $\R^+$,  we decided to use the integrals of RBF. Namely, we want to find an approximant of the form
$$ s_f(t)= \int_0^t \sum_{i=1}^M \alpha_i \phi(\lambda_i \|x - x_i\|)\ dx = \sum_{i=1}^M \alpha_i \int_0^t \phi(\lambda_i \|x - x_i\|)\ dx $$
where $\lambda_i$ is a shape parameter for every center $x_i$. As radial basis functions, we choose Gaussian functions for analytical tractability.

Evidently, any supply curve and any demand curve can be approximated by a combination of error functions, which is  the integral of a normalized Gaussian function.  The standard error function is defined as:

$${\displaystyle {\begin{aligned}\operatorname {erf} (x)={\frac {1}{\sqrt {\pi }}}\int _{-x}^{x}e^{-t^{2}}\,dt={\frac {2}{\sqrt {\pi }}}\int _{0}^{x}e^{-t^{2}}\,dt.\end{aligned}}} $$

In order to find unknown coefficients $\alpha_i, \lambda_i, x_i$ we need to solve global minimization problem: 
$$\underset{p}{\min} \|s_f(x_i,p)-y_i\|_2^2,$$ 
where $p = (\alpha_i,\lambda_i,x_i)_{i=1,\ldots,N}$ and
$$ s_f(t,p) := \sum_{i=1}^M \alpha_i \int_0^t \phi(\lambda_i \|x - x_i\|)\ dx $$
and $\phi(t) = (\mathrm{erf}(t) + 1)/2$ is the primitive of a Gaussian kernel. However, this optimization problem is very heavy, as it is a nonlinear and nonconvex minimization over $p \in \R^{3M}$. 



For this reason, we divide our global problem in simpler subproblems, with lower dimensionality, so that the final result is faster. We describe two realization of this approach in Section~\ref{sec4}. 

\section{Data set}
We now use the data about supply and demand bids from the
Italian day-ahead electricity market  from the GME website www.mercatoelettrico.org. We consider time period from 01.01 to 31.12.2017. These data are in aggregated form, i.e. bids coming from different agents, but with the same price, are aggregated in the price layer. Even in this form, we are dealing with a massive amount of data. For instance, \textbf{2 800 687} offer and \textbf{558 926} bid layers were observed during this period.

\begin{table}[h!]
\centering
\caption{Data}
\label{tab3}
\begin{tabular}{|l|c|c|c|}
\hline
Date               & \multicolumn{1}{l|}{Hour}   & \multicolumn{1}{l|}{Volume (MW)} & \multicolumn{1}{l|}{Price (Euro)} \\ \hline
01-01-2017         & 1                           & 13392.7                          & 0                                 \\ \hline
01-01-2017         & 1                           & 25                               & 0.1                               \\ \hline
01-01-2017         & 1                           & 113.8                            & 1                                 \\ \hline
01-01-2017         & 1                           & 11                               & 3.5                               \\ \hline
01-01-2017         & 1                           & 270.3                            & 5                                 \\ \hline
01-01-2017         & 1                           & 0.5                              & 6                                 \\ \hline
.................. & \multicolumn{1}{l|}{......} & ......................           & ....................              \\ \hline
31-12-2017         & 24                          & 370                              & 554.2                             \\ \hline
31-12-2017         & 24                          & 352                              & 554.3                             \\ \hline
31-12-2017         & 24                          & 365                              & 554.5                             \\ \hline
31-12-2017         & 24                          & 97                               & 700                               \\ \hline
31-12-2017         & 24                          & 60000                            & 3000                              \\ \hline
\end{tabular}
\end{table}

This means, that on average there are 324 offer and 65 bid layers for each hour of the year, which corresponds to one supply curve and one demand curve respectively. 

It is a known fact that the dynamics of electricity trade displays a set of characteristics:  external weather conditions, dependence of the consumption on the hour of the day, the day of the week, and time of the year. Variation in prices are all dependent on the principles of demand and supply. First of all, on the day-ahead market the energy is typically traded on an hourly basis and this means that the prices can and will vary per hour. For example, at 9:00 a.m. there could be a price peak, while at 4:00 a.m. prices could be only half of the peak price. Second, the weekly seasonal behaviour matters. Usually, it is necessary to differentiate between the two weekend days (Saturday and Sunday), the first business day of the week (Monday), the last business day of the week (Friday) and the remaining business days (see e.g. \cite{Andreis}). Thirdly, electricity spot prices display a strong yearly seasonal pattern: for instance, demand increases in summer, as consumers turn  their air conditioners on, and also in winter because of electric heating in housing.

\begin{figure}[H]
\begin{minipage}{0.5\textwidth}
 \includegraphics[width=1\textwidth]{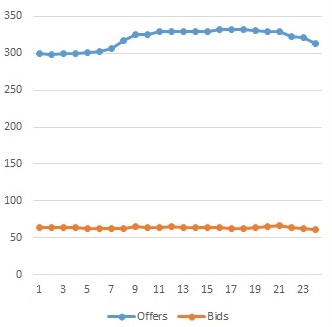}    
   \end{minipage}
  \begin{minipage}{0.5\textwidth} 
\begin{tabular}{|c|c|c|c|c|c|}
\hline
\multicolumn{1}{|l|}{\multirow{2}{*}{Hour}} & \multicolumn{2}{c|}{Number}                                   & \multicolumn{1}{l|}{\multirow{2}{*}{Hour}} & \multicolumn{2}{c|}{Number}                                   \\ \cline{2-3} \cline{5-6} 
\multicolumn{1}{|l|}{}                      & \multicolumn{1}{l|}{of offers} & \multicolumn{1}{l|}{of bids} & \multicolumn{1}{l|}{}                      & \multicolumn{1}{l|}{of offers} & \multicolumn{1}{l|}{of bids} \\ \hline
\textbf{1}                                  & 300                            & 64                           & \textbf{13}                                & 329                            & 64                           \\ \hline
\textbf{2}                                  & 299                            & 64                           & \textbf{14}                                & 329                            & 64                           \\ \hline
\textbf{3}                                  & 300                            & 64                           & \textbf{15}                                & 330                            & 64                           \\ \hline
\textbf{4}                                  & 300                            & 64                           & \textbf{16}                                & 332                            & 64                           \\ \hline
\textbf{5}                                  & 301                            & 63                           & \textbf{17}                                & 332                            & 63                           \\ \hline
\textbf{6}                                  & 303                            & 63                           & \textbf{18}                                & 332                            & 63                           \\ \hline
\textbf{7}                                  & 307                            & 62                           & \textbf{19}                                & 331                            & 64                           \\ \hline
\textbf{8}                                  & 318                            & 63                           & \textbf{20}                                & 329                            & 65                           \\ \hline
\textbf{9}                                  & 325                            & 65                           & \textbf{21}                                & 329                            & 66                           \\ \hline
\textbf{10}                                 & 326                            & 64                           & \textbf{22}                                & 323                            & 64                           \\ \hline
\textbf{11}                                 & 329                            & 64                           & \textbf{23}                                & 321                            & 63                           \\ \hline
\textbf{12}                                 & 329                            & 65                           & \textbf{24}                                & 314                            & 61                           \\ \hline
\end{tabular}
   \end{minipage} 
\caption{Hour dependence of the number of offer and bid layers}\label{off_dep_hour}
\end{figure}

\begin{figure}[H]
     \begin{minipage}{0.5\textwidth}
    \includegraphics[width=1\textwidth]{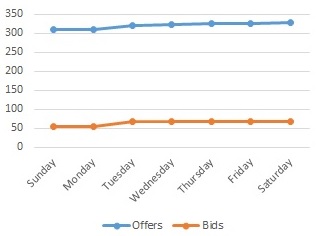}   
   \end{minipage}
   \begin{minipage}{0.4\textwidth}
\begin{tabular}{ccclll}
\cline{1-3}
\multicolumn{1}{|c|}{\multirow{2}{*}{Month}} & \multicolumn{2}{c|}{Number}                                   & \multicolumn{3}{l}{\multirow{7}{*}{}}          \\ \cline{2-3}
\multicolumn{1}{|c|}{}                       & \multicolumn{1}{l|}{of offers} & \multicolumn{1}{l|}{of bids} & \multicolumn{3}{l}{}                           \\ \cline{1-3}
\multicolumn{1}{|c|}{\textbf{Sunday}}        & \multicolumn{1}{c|}{310}       & \multicolumn{1}{c|}{55}      & \multicolumn{3}{l}{}                           \\ \cline{1-3}
\multicolumn{1}{|c|}{\textbf{Monday}}        & \multicolumn{1}{c|}{310}       & \multicolumn{1}{c|}{56}      & \multicolumn{3}{l}{}                           \\ \cline{1-3}
\multicolumn{1}{|c|}{\textbf{Tuesday}}       & \multicolumn{1}{c|}{322}       & \multicolumn{1}{c|}{68}      & \multicolumn{3}{l}{}                           \\ \cline{1-3}
\multicolumn{1}{|c|}{\textbf{Wednesday}}     & \multicolumn{1}{c|}{324}       & \multicolumn{1}{c|}{67}      & \multicolumn{3}{l}{}                           \\ \cline{1-3}
\multicolumn{1}{|c|}{\textbf{Thursday}}      & \multicolumn{1}{c|}{326}       & \multicolumn{1}{c|}{68}      & \multicolumn{3}{l}{}                           \\ \cline{1-3}
\multicolumn{1}{|c|}{\textbf{Friday}}        & \multicolumn{1}{c|}{327}       & \multicolumn{1}{c|}{68}      & \multicolumn{3}{c}{\multirow{5}{*}{\textbf{}}} \\ \cline{1-3}
\multicolumn{1}{|c|}{\textbf{Saturday}}      & \multicolumn{1}{c|}{329}       & \multicolumn{1}{c|}{68}      & \multicolumn{3}{c}{}                           \\ \cline{1-3}
                        
\end{tabular}
   \end{minipage} 
    \caption{Weekly dependence of the number of offer and bid layers}\label{off_dep_week}
\end{figure}

\begin{figure}[H]
      \begin{minipage}{0.6\textwidth}
     \includegraphics[width=1\textwidth]{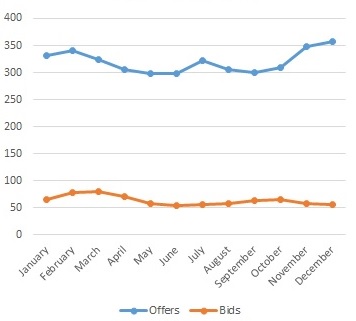}    
   \end{minipage}
   \begin{minipage}{0.5\textwidth}
\begin{tabular}{|c|c|c|ccc}
\cline{1-3}
\multirow{2}{*}{Month} & \multicolumn{2}{c|}{Number}                                   & \multicolumn{1}{l}{\multirow{2}{*}{}} & \multicolumn{2}{c}{}                        \\ \cline{2-3}
                       & \multicolumn{1}{l|}{of offers} & \multicolumn{1}{l|}{of bids} & \multicolumn{1}{l}{}                  & \multicolumn{1}{l}{} & \multicolumn{1}{l}{} \\ \cline{1-3}
\textbf{January}       & 331                            & 65                           & \textbf{}                             &                      &                      \\ \cline{1-3}
\textbf{February}      & 341                            & 79                           & \textbf{}                             &                      &                      \\ \cline{1-3}
\textbf{March}         & 324                            & 81                           & \textbf{}                             &                      &                      \\ \cline{1-3}
\textbf{April}         & 305                            & 72                           & \textbf{}                             &                      &                      \\ \cline{1-3}
\textbf{May}           & 298                            & 57                           & \textbf{}                             &                      &                      \\ \cline{1-3}
\textbf{June}          & 298                            & 54                           & \textbf{}                             &                      &                      \\ \cline{1-3}
\textbf{July}          & 322                            & 55                           & \textbf{}                             &                      &                      \\ \cline{1-3}
\textbf{August}        & 305                            & 58                           & \textbf{}                             &                      &                      \\ \cline{1-3}
\textbf{September}     & 300                            & 64                           & \textbf{}                             &                      &                      \\ \cline{1-3}
\textbf{October}       & 309                            & 66                           & \textbf{}                             &                      &                      \\ \cline{1-3}
\textbf{November}      & 348                            & 58                           & \textbf{}                             &                      &                      \\ \cline{1-3}
\textbf{December}      & 357                            & 57                           & \textbf{}                             &                      &                      \\ \cline{1-3}
\end{tabular}
   \end{minipage} 

    \caption{Monthly dependence of the number of offer and bid layers}\label{off_dep_mon}
\end{figure}

As far as the number of offers (or bids) affects directly the complexity of approximation, we decided to explore the relationship between the number of bids and offers and such a characteristics as the hour of the day, the day of the week, and the month of the year. Based on the dependence between this three factors and electricity prices we could expect that some hours, days have much less offers and bids than another one. This analysis is presented on Figures~\ref{off_dep_hour}~--~\ref{off_dep_mon}.

The main conclusion that we have made is that there is no direct relationship between the number of offer and bid layers and the hour of the day, the day of the week, and the time of the year. In particular, during 24 hour of the day the number of offer layers varies between 299 and 332, and the number of bid layers varies between 61 and 66. With regard to dependence of the day of the week the number of offer layers varies between 310 and 320, and the number of bid layers varies between 55 and 68. Based on this observation we decided to choose the same number of basis functions independently of the hour of the day, the day of the week, and the time of the year.

\section{Numerical experiments}\label{sec4}

Since the maximum  market clearing price for the period under review (i.e. from 01.01.2017 to 31.12.2017) is 350 \euro, in all the experiments we restricted ourselves to a maximum price of 400 \euro. For the realization of our algorithm we are using the  function \texttt{lsqcurvefit} from  MATLAB Optimization Toolbox.

First, we download the data from a text file and  choose the number of basis function $M$. After that, we need to divide our problem into $M$ sub-problems. Then each part of the supply curve must be approximated by one error function. 

\begin{figure}[h!]
\begin{minipage}[b]{.4\textwidth}
\centering
\includegraphics[width=1\textwidth]{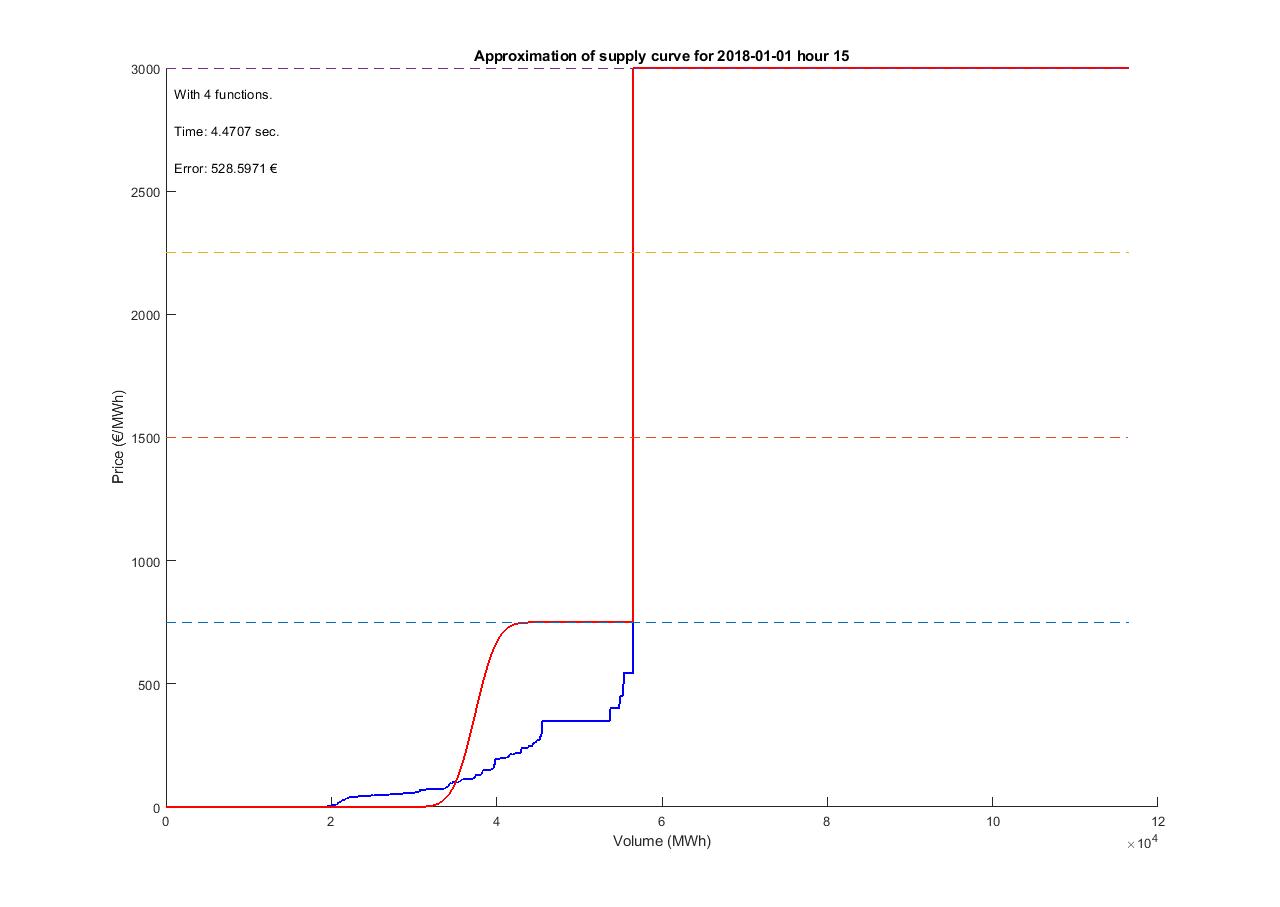}
\caption{Method 1}\label{methods1}
\end{minipage}
\hfill
\begin{minipage}[b]{.4\textwidth}
\centering
\includegraphics[width=1\textwidth]{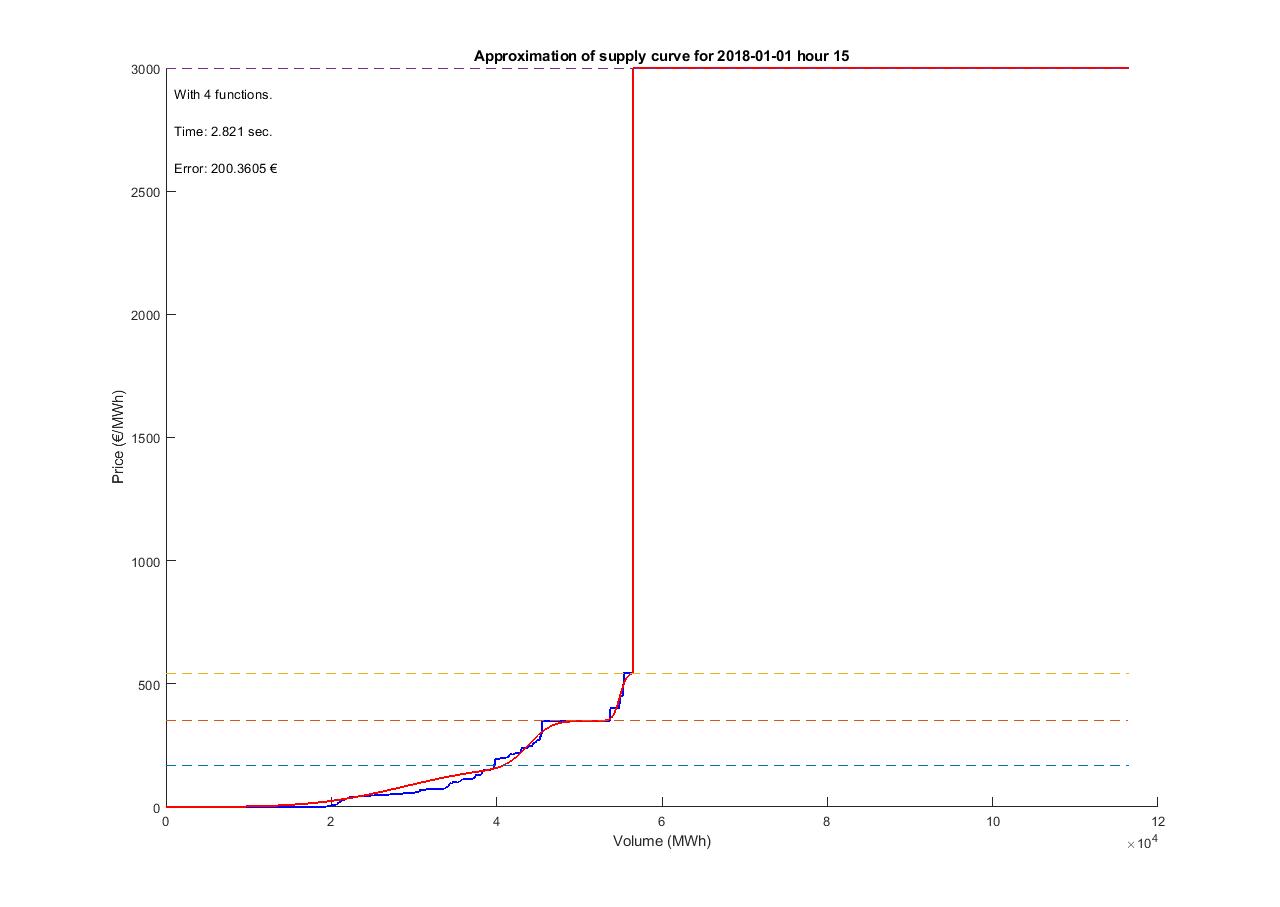}
\caption{Method 2} \label{methods2}
\end{minipage}
\end{figure}

Our first attempt (Method 1) was just to divide $y$-axis uniformly into $M$ equal intervals (see  Figure \ref{methods1}). However this approach is ineffective, as a huge jump concentrates on itself, keeping uselessly many components. 
 
To resolve this problem we created a simple algorithm - Method 2 - that finds the points $p_1, \ldots, p_M$ on the $y$-axis such that our supply curve takes the value exactly $p_i$ on some non-trivial interval (see  Figure \ref{methods2}).  

\begin{figure}[H]
\caption{Local interpolation by one error function with \texttt{lsqcurvefit} function}
\label{Devided}
\centering
\includegraphics[width=0.45\linewidth]{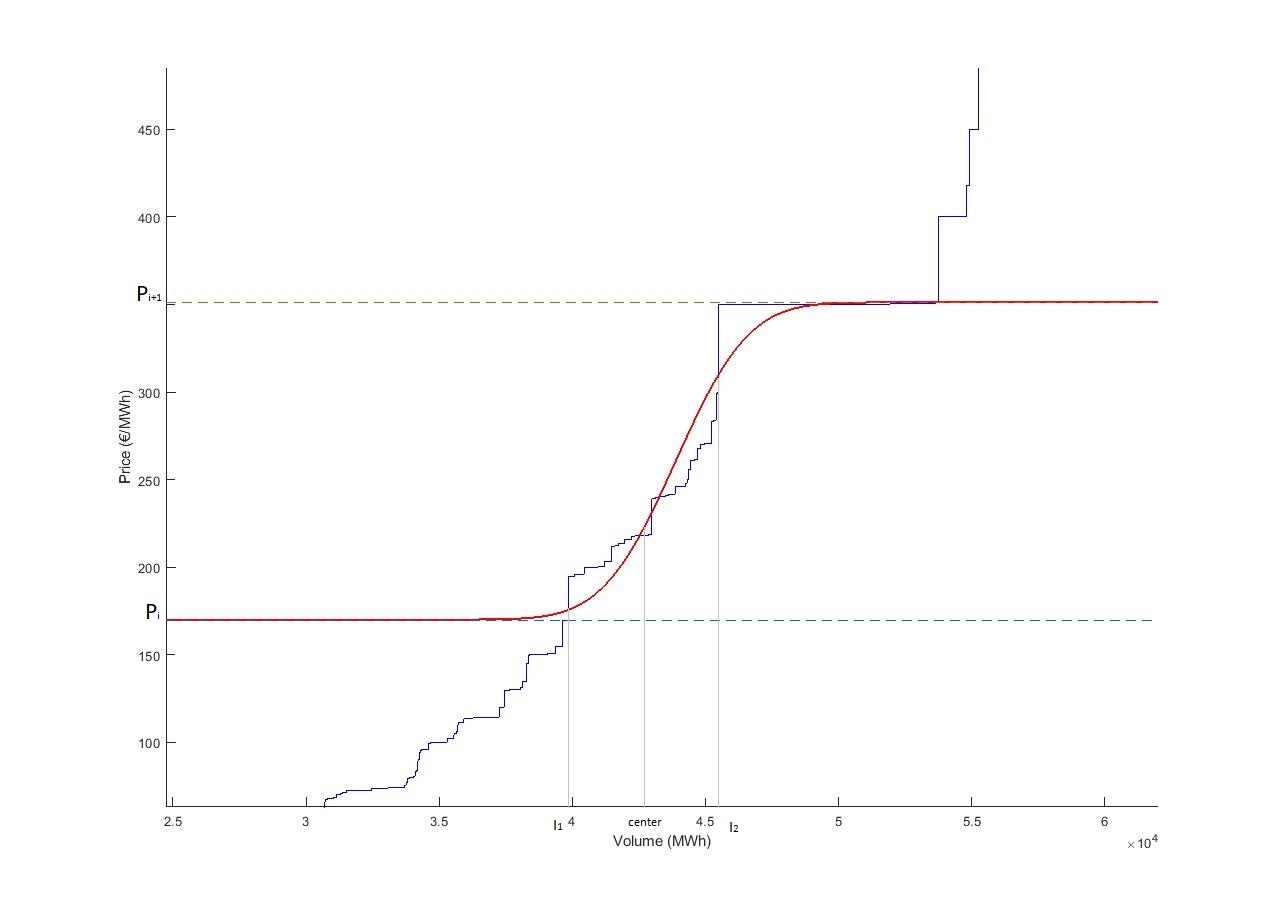}  
\includegraphics[width=0.45\linewidth]{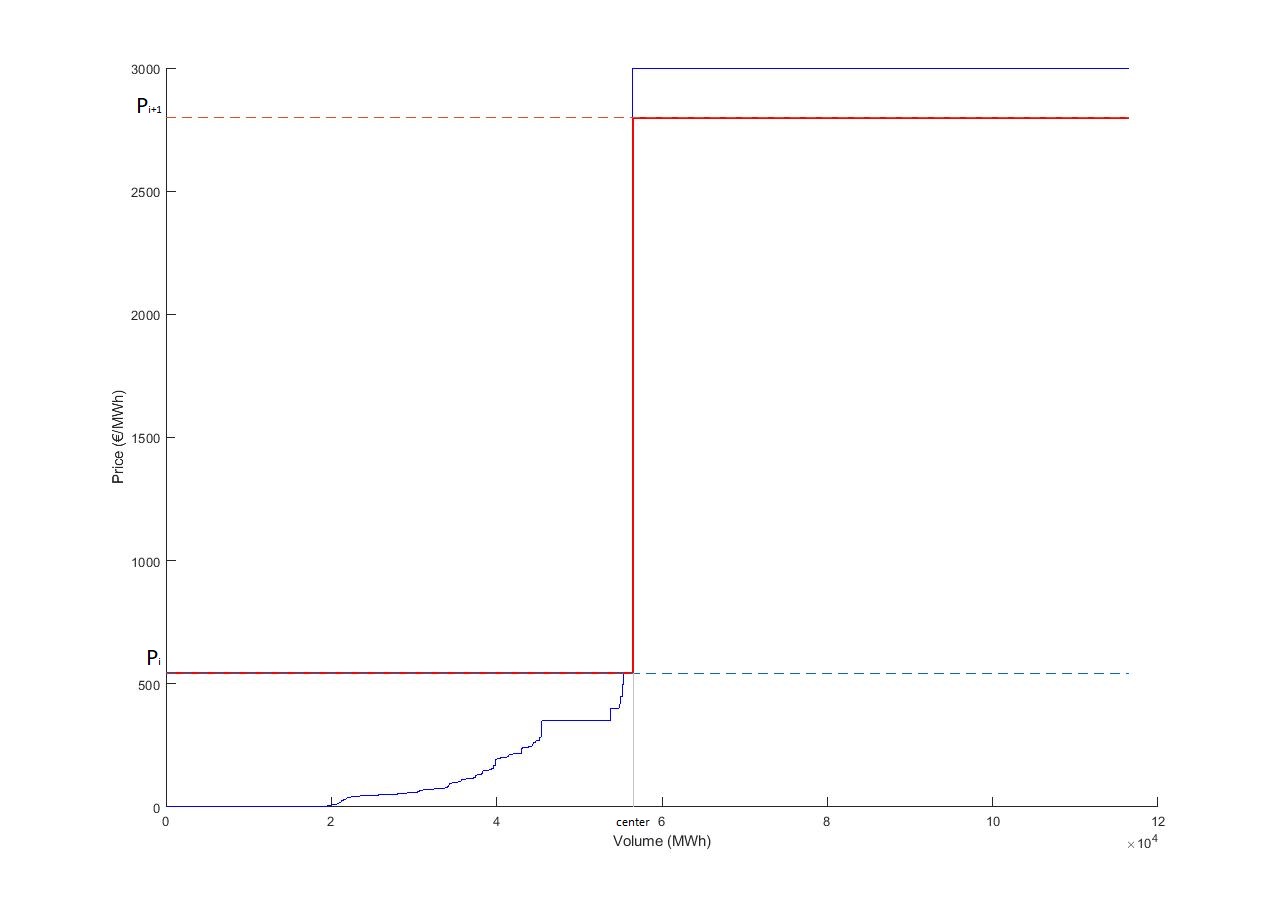}
\label{Algorithm}
\end{figure}

Then  $M$ times we resolve the same optimization problem for the values of the supply curve between $p_i$ and $p_{i+1}$ using function \texttt{lsqcurvefit} (see Figure \ref{Devided}). On each part we need to find only 3 coefficients $a_i, b_i, c_i$ of the function \begin{equation}
G(x)=\sum_{i=1}^k a_i (\erf(c_i\cdot(x-b_i))+1).
\end{equation}
 Here, for convenience of representation we are using $\{\erf(c_i\cdot(x-b_i))+1\}$ instead of $\{\erf(c_i\cdot(x-b_i))\}$, as our data values are never negative.

The \texttt{lsqcurvefit} function solves nonlinear data-fitting problems in least-squares sense. Suppose that we have   data points  $X_N=\{x_i: i=1, 2, \ldots, N\}$  and  data values $Y_N=\{y_i: i=1, 2,\ldots, N\}\subset \R$ and we want to find a function $f$ such that $f(x_i)\approx y_i,\, i=1,\ldots,N.$ We can consider the family of functions $\{f(x,p): p\in \R^k\}$, depending of some parameter $p\in \R^k$. Let $p_0\in \R^k$ be an ``initial guess'' such that $f(x_i,p)$ is reasonably close to $y_i$. The function \texttt{lsqcurvefit} starts at $p_0$ and finds coefficients $p$ from some neighborhood of $p_0$ to best fit the data set $Y_N$:
$$\underset{p}{\min}\|f(x_i,p)-y_i\|_2^2.$$ 

Notice that this function works well only if the number of parameters $(p_1, \ldots, p_k)$ is not very big. That is why we are forced  to divide our problem into many local problems.

For optimizing the numerical procedure we solved some parts of the optimization problem by ourselves: in fact, when the interval $[p_i,p_{i+1}]$ contains only one jump, then
$$ a_i := f(p_{i+1}) - f(p_i) $$
for any kernel function $\phi$ with unit integral.


A summary of the results is shown in Table \ref{tab2}. For all experiments we proceed with the data for period from 01.01.2017 to 31.12.2017. We used different number of basis function to approximate supply and demand curves, and then compared the equilibrium price, which was received as intersection of approximants ($P_{appr}$), with the correct equilibrium price ($P$). We did this for each hour of each day, and then computed the average value of $|P - P_{appr}|$ (Error) for all 8 664 hours of the year and the maximum value of $|P - P_{appr}|$ (Max error).

This empirical results show that the accuracy of our approximation is good enough, if we use 5 basis function for the demand curve and 15 basis function for the supply curve. Then the increase in the number of functions leads to more time consumption, but the increase of the accuracy is less significant.

\bigskip
\begin{table}[H]
\centering
\caption{Results of numerical experiment}
\label{tab2}
\begin{tabular}{|c|c|c|c|c|} 
\hline
\multicolumn{2}{|c|}{Number of functions}  & \multicolumn{3}{c|}{Results} \\ 
\hline
For demand & For supply & Error & Max error &Running time \\
\hline
5   &5   & 3.9 \euro  &28.6 \euro & 69 min.     \\ 
5   &10   & 2.2 \euro & 14.9 \euro & 82 min.     \\ 
5   &15  & 1.5  \euro & 11.1 \euro & 103 min.      \\ 
5   &20  & 1.3   \euro & 9.1 \euro & 110 min.     \\ 
5   &25  & 1.2  \euro & 9.3 \euro & 135 min.      \\ 
5   &30 & 1.2  \euro & 9.4 \euro & 159 min.   \\ 
5   &35 & 1.2  \euro & 9.8 \euro & 177 min.     \\ 
5   &40 & 1.2   \euro & 9.6 \euro & 190 min.     \\ 
5   &45 & 1.2   \euro & 9.6 \euro & 199 min.     \\ 
5   &50 & 1.2   \euro & 9.6 \euro & 207 min.     \\ 
\hline
10   &5   & 3.9 \euro & 39.5 \euro & 100 min.     \\ 
10 &10   & 2.1 \euro & 14.9 \euro & 128 min.     \\ 
10   &15  & 1.4  \euro & 8.9 \euro  & 146 min.     \\ 
10   &20  & 1.2    \euro & 9.1 \euro  & 162 min.     \\ 
10    &25  &  1.1 \euro & 9.5 \euro & 183 min.     \\ 
10    &30 & 1.1 \euro & 9.3 \euro & 199 min.     \\ 
10    &35 & 1.0 \euro & 9.4 \euro & 223 min.     \\ 
10    &40 & 0.98  \euro & 9.8 \euro & 241 min.     \\ 
10    &45 & 0.98   \euro & 9.6 \euro & 255 min.     \\ 
10    &50 & 0.98   \euro & 9.6 \euro & 273 min.     \\ 
\hline
\end{tabular}
\end{table}

As a last step we analyzed  the stability of the coefficients for the case when we approximate the supply curve with 10 basis functions and the demand curve with 5 basis functions for the same period of time, as
$$S(x)=\sum_{i=1}^{10} A_i (\erf(C_i\cdot(x-B_i))+1)\quad\text{ and } \quad D(x)=\sum_{i=1}^5 E_i (\erf(K_i\cdot(x-L_i))+1).$$

From Table \ref{tab3} we can see that these coefficients do not have a stable behavior (namely, maximum values, minimum values and mean values are presented). Although the values attained by the supply curve are clustered around  layers, which correspond to different 
production technologies, we came to the conclusion that we have no chance to choose these coefficients uniformly for all curves, but we need to calculate them for all supply and demand curves.
\begin{figure}[H]
\caption{Supply curve approximated with 10 basis functions} 
\centering
\includegraphics[width=1\linewidth]{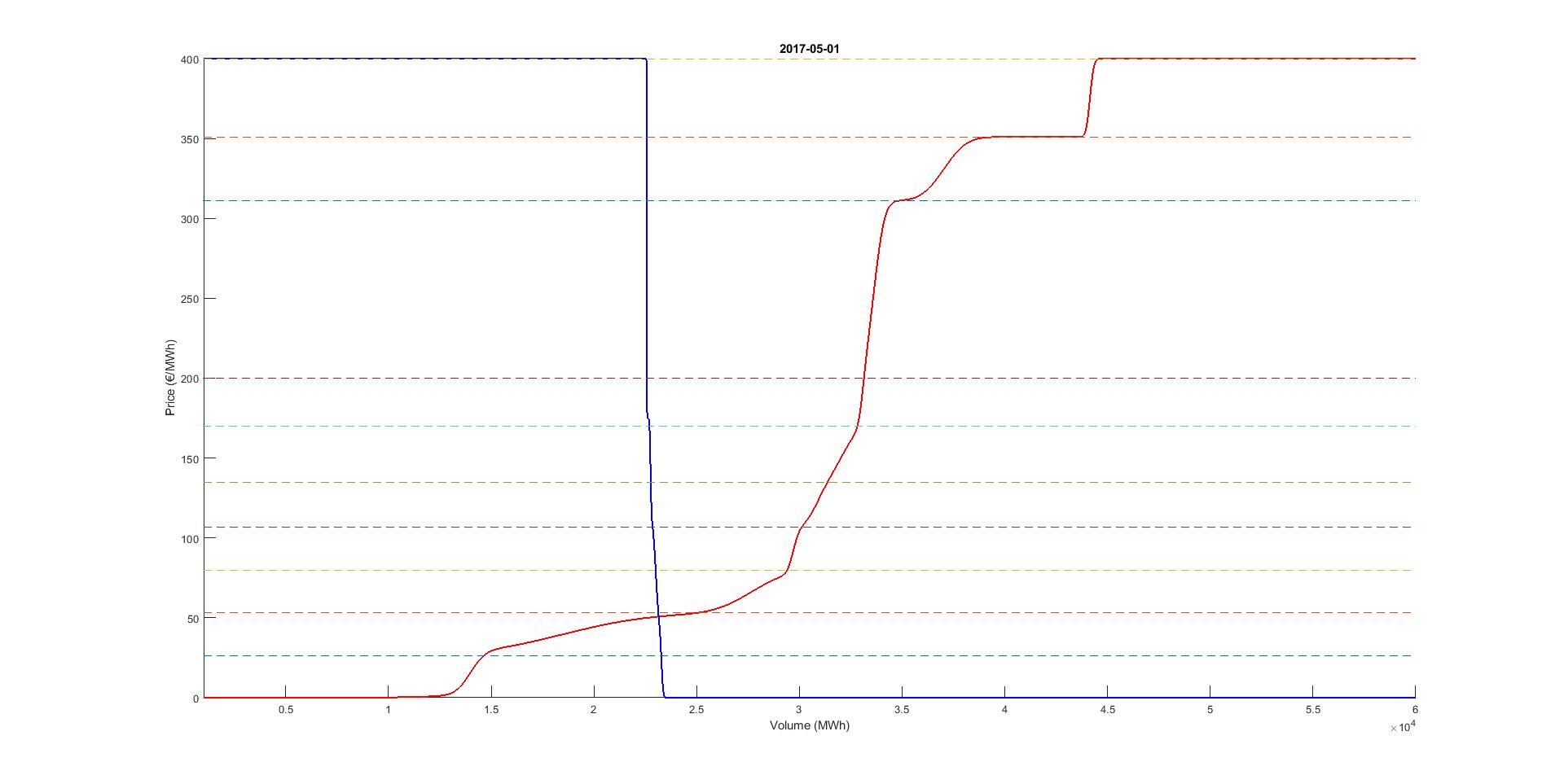}  
\label{Algorithm}
\end{figure}
\begin{table}[H]
\centering
\caption{Stability of the coefficients}
\label{tab3}
\begin{tabular}{|c|c|c|c|}
\hline
    &  Min                          & Mean     & Max  \\
    \hline
    \multicolumn{4}{|c|}{Coefficients for supply curve}  \\
    \cline{1-4}
$A_1$  & 10                           & 14.76981 & 18    \\
$A_2$  & 10.5                           & 15.15519 &21   \\
$A_3$  &   10.5                       & 15.21438 & 19.5  \\
$A_4$  & 11                             & 15.53944 & 22   \\
$A_5$  & 11                          & 16.8968  & 27.5    \\
$A_6$  & 12.5                             & 20.44287 & 27 \\
$A_7$  & 14.5                            & 22.15457 & 33  \\
$A_8$  &19                         & 29.69132 & 57.5     \\
$A_9$  &17                              & 24.48784 & 48   \\
$A_{10}$ & 21                             & 25.64777 & 50   \\
\hline
\multicolumn{4}{|c|}{Coefficients for demand curve}    \\
\cline{1-4}
$E_1$  & 12                          & 30.95154 & 37.5    \\
$E_2$  & 25                           & 34.31039 & 58.5   \\
$E_3$  & 25                             & 36.24469 & 50   \\
$E_4$ & 33                            & 40.19715 & 50    \\
$E_5$  & 50                             & 58.29623 & 75  \\
\hline
\end{tabular}
\end{table}

\section{Conclusions}

 We presented a parsimonious way to represent supply and demand curves, using a mesh-free method based on Radial Basis Functions.  Using the tools of functional data analysis, we are able to approximate the original curves with far less parameters  than the original ones.  Namely, in order to approximate piece-wise constant monotone  functions,  we are  using linear combinations of integrals of Gaussian functions. 

The real data about supply and demand bids from the Italian day-ahead electricity market showed that there is no direct relationship between the number of offer and bid layers and the hour of the day, the day of the week, and the time of the year. Based on this observation, we decided to choose the same number of basis functions independently of the hour of the day, the day of the week, and the time of the year.

The numerical results showed that the accuracy of our approximation is good enough, if we use 5 basis function for the demand curve and 15 basis function for the supply curve, and then the increase in the number of functions leads to more time-consumption, but the increase of the accuracy is less significant.

\section*{Acknowledgements}

The authors thank Enrico Edoli, Marco Gallana, and Emma Perracchione for several useful discussions.
The authors wish to thank also Florian Ziel, Carlo Lucheroni, Stefano Marmi, Sergei Kulakov, Enrico Moretto for their comments and suggestions. The authors would like to thank the participants of the following events: Energy Finance Christmas Workshop (2018) in Bolzano; Quantitative Finance Workshop (2019) in Zurich; Energy Finance Italia Workshop in Milan (2019), the Freiburg-Wien-Zurich Workshop (2019) in Padova.

The first author is pursuing her Ph.D. with a fellowship for international students funded by Fondazione Cassa di Risparmio di Padova e Rovigo (CARIPARO) and acknowledges the support of this project. The second author acknowledges financial support from the research projects of the University of Padova BIRD172407-2017 "New perspectives in stochastic methodsfor finance and energy markets" and BIRD190200/19 "Term Structure Dynamics in Interest Rate  and  Energy  Markets:  Modelling  and  Numerics".


\begin{thebibliography}{99}
\bibitem{Andreis}
Andreis, L., Flora, M., Fontini, F. and Vargiolu, T. (2019) Pricing Reliability Options under different electricity prices' regimes, preprint, https://arxiv.org/abs/1909.05761.
\bibitem{Aneiros}
Aneiros G., Vilar J.M., Cao R., San Roque A.M. Functional prediction for the
residual demand in electricity spot markets. IEEE Trans. Power Syst. 28 (4) (2013), 4201-4208.

\bibitem{Aronszajn} Aronszajn N.,  Theory of Reproducing Kernels, Transactions of the American Mathematical Society. 68 (3) (1950), 337-404. 

\bibitem{Bozzini} Bozzini, M. Lenarduzzi, L., Rossini M., Schaback R.  Interpolation with variably scaled kernels. IMA Journal of Numerical Analysis, 35(1) (2015), 199-219.

\bibitem{Eichler} Eichler, M., Sollie, J., Tuerk, D.  A new approach for modelling electricity spot
prices based on supply and demand spreads. Conference on Energy Finance 2012
Trondheim, Norway. pp. 1-4.

\bibitem{Fasshauer}
Fasshauer G.E.,  McCourt M.J. Kernel-based Approximation Methods Using
MATLAB, World Scientific, Singapore, 2015.

\bibitem{Fasshauer2011}  Fasshauer, G.E. Positive definite kernels: past, present and future,
Dolomite Research Notes on Approximation 4 (2011), 21-63

\bibitem{Mercer}
J. Mercer, Functions of positive and negative type and their connection with the
theory of integral equations, Phil. Trans. Royal Society 209 (1909), 415-446.

\bibitem{Micchelli}
 Micchelli C.A. Interpolation of scattered data: Distance matrices and conditionally positive definite functions, Constr. Approx. 2 (1986), 11-22.

\bibitem{Weron2} Nowotarski  J.,  Weron  R.   Recent  advances  in  electricity  price  forecasting:  A  review  of  probabilisticforecasting. Renew. Sustain. Energy Rev., 81 (2018), 1548-1568. 

















\bibitem{Emma1} Perracchione E.,  Stura I. RBF kernel method and its applications to clinical data, Dolomites Res. Notes Approx. 9 (2016),  13-18.

\bibitem{Schaback} Schaback R.  Native Hilbert Spaces for Radial Basis Functions I. In: M\"uller M.W., Buhmann M.D., Mache D.H., Felten M. (eds) New Developments in Approximation Theory. ISNM International Series of Numerical Mathematics, vol 132. Birkh\"auser, Basel 

\bibitem{Schaback2} Schaback R. A unified theory of radial basis functions: Native Hilbert spaces for radial basis functions II. Journal of computational and applied mathematics 121 (1-2) (2000), 165-177.

\bibitem{Tesi}
Shah, I.  Modeling and Forecasting Electricity Market Variables (2016) [PhD Thesis] .





 











\bibitem{Ugurlu}  Ugurlu U., Oksuz I., Tas O. Electricity Price Forecasting Using Recurrent Neural Networks. Energies 11(5), 1255 (2018), 1255.

\bibitem{Wendland}
 Wendland H. Scattered Data Approximation, Cambridge Monogr. Appl. Comput.
Math., vol. 17, Cambridge Univ. Press, Cambridge, 2005.

\bibitem{Weron}  Weron R. Electricity price forecasting: A review of the state-of-the-art with a look into the future, Int. J. Forecast, 30(4) (2014), 1030 - 1081.

\bibitem{Ziel1}  Ziel F., Steinert R.
Electricity price forecasting using sale and purchase curves: the X-Model, Energy Econ., 59 (2016), 435-454.
\end{thebibliography}
\end{document}